\documentclass[12pt]{article}
\usepackage{pdproc,epsfig} 

\begin{document}

\newcommand{\lwig}{\mbox{\,\raisebox{.3ex}
    {$<$}$\!\!\!\!\!$\raisebox{-.9ex}{$\sim$}\,}}
\newcommand{\gwig}{\mbox{\,\raisebox{.3ex}
    {$>$}$\!\!\!\!\!$\raisebox{-.9ex}{$\sim$}}\,}
\newcommand{\lambdabar}{{\hbox{$\lambda_e$\kern-1.9ex\raise+0.45ex\hbox{--}
\kern+0.2ex}}}

\title{
{\normalsize\rm \rightline{DESY 07-054}}
\vskip 1cm 
 PARTICLE INTERPRETATIONS OF THE PVLAS DATA\footnote{Talk presented at the 
XII International Workshop on ``Neutrino Telescopes'', March 6-9, 2007, Venice, Italy.}
}

\author{ANDREAS RINGWALD}

\address{ Deutsches Elektronen-Synchrotron DESY, Notkestra\ss e 85\\
 D-22607 Hamburg, Germany\\
 {\rm E-mail: andreas.ringwald@desy.de}}

\abstract{Recently the PVLAS collaboration reported the observation of a
rotation of linearly polarized laser light induced by a transverse
magnetic field -- a signal being unexpected within standard QED. In this
review, we emphasize two mechanisms which have been proposed to explain 
this result: production of a single light neutral spin-zero particle or pair
production of light minicharged particles. 
We discuss a class of models, involving, in addition to our familiar ``visible'' photon, 
further light ``hidden paraphotons'', which mix kinematically with the visible one,  
and further light paracharged particles. In these
models, very strong astrophysical and cosmological bounds on the weakly interacting light particles
mentioned above can be evaded. In the upcoming year,
a number of decisive laboratory based tests of the particle interpretation of
the PVLAS anomaly will be done. More generally, such experiments, 
exploiting high fluxes of low-energy photons and/or large electromagnetic fields, 
will dig into
previously unconstrained parameter space of the above mentioned models.
}
   
\normalsize\baselineskip=15pt

\section{Introduction}

We are entering a new era in particle physics: Next year, the Large Hadron Collider 
(LHC) will start to probe, through the collision of $7$~TeV protons, the structure of
matter and space-time at an unprecedented level. There is a lot of circumstantial
evidence that the physics at the TeV scale exploited at LHC and later at the 
International Linear Collider (ILC) will bring decisive insights into fundamental
questions such as the origin of particle masses, the nature of dark matter in the universe, and
the unification of all forces, including gravity. 
Indeed, most proposals to embed the standard model of particle physics into a more general, 
unified framework, notably the ones based on string theory or its low energy incarnations, 
supergravity and supersymmetry, predict new heavy, $m\gg 100$~GeV, 
particles which may be searched for at TeV colliders. Some of these particles, 
prominent examples being neutralinos, are natural candidates for the constituents of 
cold dark matter in the form of so-called weakly interacting massive particles (WIMPs).   

However, there is also evidence that there is fundamental physics at the sub-eV scale. 
Indeed, atmospheric, reactor, and solar neutrino data strongly support the hypothesis
that neutrinos have masses in the sub-eV range. Moreover, the vacuum energy density of
the universe, as inferred from cosmological observations, points to the sub-eV range, 
$\rho_\Lambda \sim {\rm meV}^4$. As a matter of fact, many of the above mentioned extensions
of the standard model not only predict WIMPs, but also WILPs, i.e. weakly interacting
light particles, some of them even having possibly a tiny electric charge 
(so-called minicharged particles). Prominent candidates for such particles go under the 
names axions, dilatons, and moduli. 
Unlike for WIMPs, TeV colliders are not the best means to search for WILPs. 
For this purpose, small, high-precision experiments, exploiting high fluxes of low-energy photons and/or
large electromagnetic fields, seem to be superior.

\begin{figure}
\begin{center}
\psfig{file=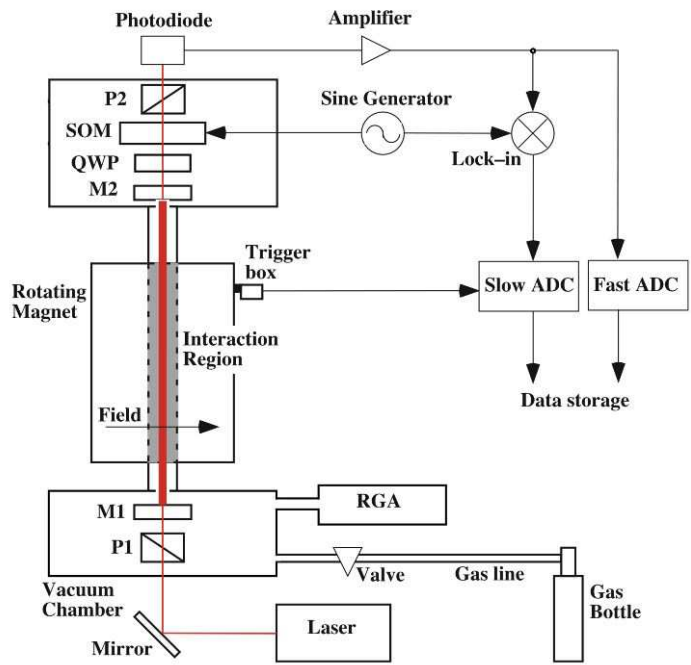,width=10cm}
\end{center}
\caption[]{Schematic illustration of the PVLAS experiment\cite{Zavattini:2005tm}.}
\label{fig:pvlas}
\end{figure}

\begin{figure}
\begin{center}
\psfig{file=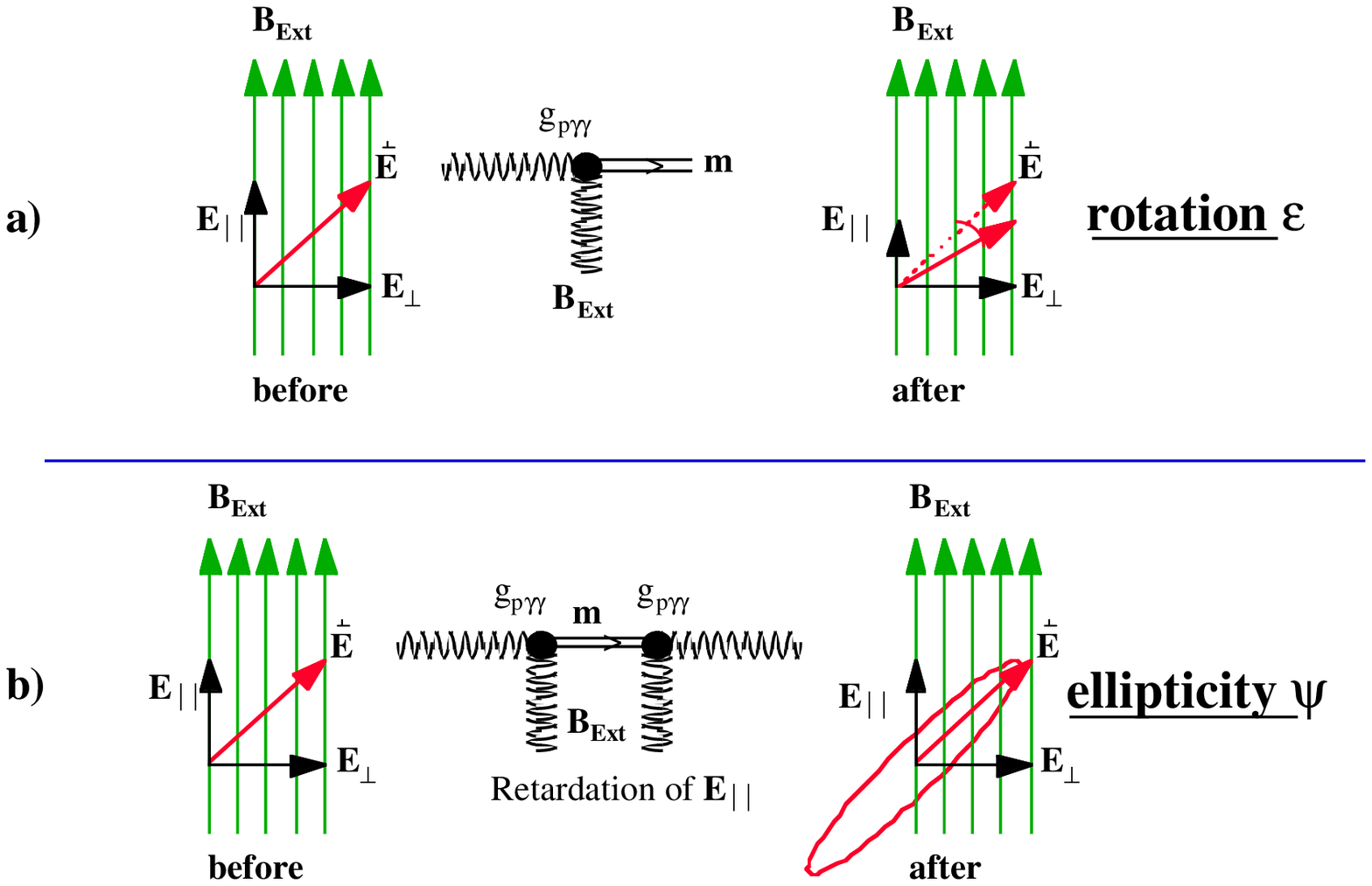,width=11cm}

\vspace{2cm}
\psfig{file=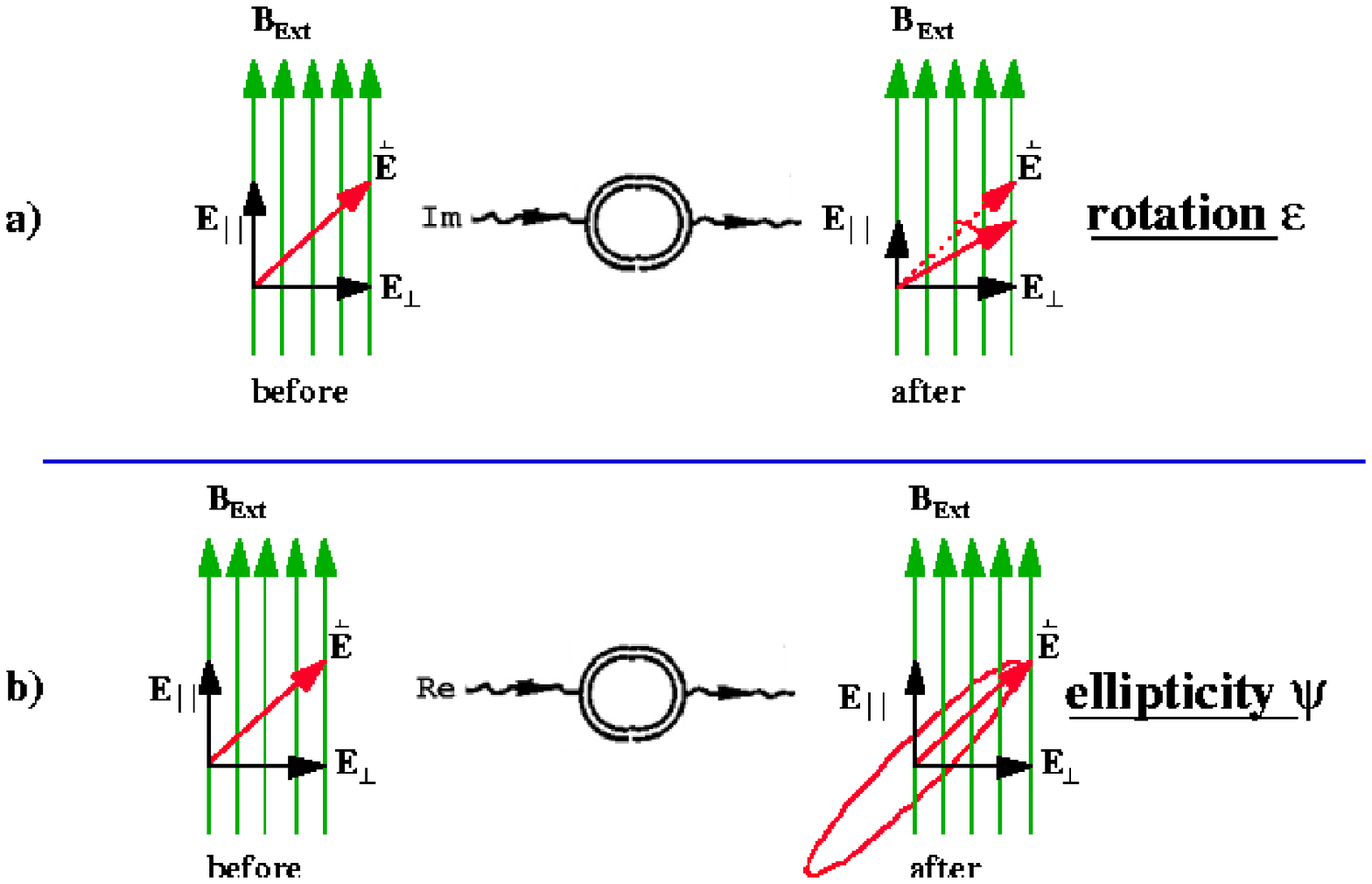,width=11cm}
\end{center}
\caption[]{Changes of the polarization state of initially linearly polarized 
photons after the passage through a magnetic field, due to real and virtual 
conversion into an axion-like particle (from Ref.~\cite{Brandi:2000ty}) (top panel) or 
into a pair of minicharged particles (bottom panel). The double lines in the bottom panel denote
the exact propagator of the minicharged particle in the background of the magnetic field.}
\label{fig:pol_int}
\end{figure}

\begin{table}
\caption[]{Current experimental data on vacuum magnetic dichroism, birefringence, and on photon 
regeneration.\\ 
{\em Top:} The vacuum rotation ${\Delta\theta}$, ellipticity $\psi$ and
photon regeneration rate from the BFRT\cite{Cameron:1993mr}
experiment. For the polarization data,
BFRT used a magnetic field with time-varying amplitude $B=B_0+\Delta
B\cos(\omega_m t+\phi_m)$, where $B_0=3.25$~T and $\Delta B=0.62$~T. 
For photon regeneration, they employed
$B=3.7$~T.\\
{\em Middle:} The vacuum rotation ${\Delta\theta}$ and ellipticity $\psi$
  per pass measured by 
PVLAS, for $B=5$~T. 
The rotation of polarized laser light
  with $\lambda=1064$~nm is published in Ref.~\cite{Zavattini:2005tm}.
  Preliminary results are taken from
  Refs.~\cite{PVLASICHEP,Cantatore:IDM2006,Zavattini:NT07} {and are used here
  for illustrative purposes only}.\\
{\em Bottom:} The vacuum rotation ${\Delta\theta}$ from the Q\&A
experiment\cite{Chen:2006cd} ($B=2.3$~T).}\label{tab:pol_data}
  \small

\vspace{3ex}
\begin{center}
\begin{tabular}{ccc}
\hline\hline
\multicolumn{3}{c}{\makebox[8cm][c]{\bf {\bf BFRT} experiment}}\\
\hline
\multicolumn{3}{c}{{\bf Rotation}\hfill($L=8.8$~m, $\lambda=514.5$~nm, $\theta=\frac{\pi}{4}$)}\\[0.1cm]
\makebox[2.5cm][c]{$N_{\rm pass}$}&\makebox[2.5cm][c]{$\left|{\Delta\theta}\right|\,[{\rm nrad}]$}&
\makebox[2.5cm][c]{${\Delta\theta}_{\rm noise}\,[{\rm nrad}]$}\\[0.1cm]
$254$&$0.35$&$0.30$\\
$34$&$0.26$&$0.11$\\
\hline
\multicolumn{3}{c}{{\bf Ellipticity}\hfill($L=8.8$~m, $\lambda=514.5$~nm, $\theta=\frac{\pi}{4}$)}\\[0.1cm]
\makebox[2.5cm][c]{$N_{\rm pass}$}&\makebox[2.5cm][c]{$\left|\psi\right|\,[{\rm nrad}]$}&
\makebox[2.5cm][c]{$\psi_{\rm noise}\,[{\rm nrad}]$}\\[0.1cm]
$578$&$40.0$&$11.0$\\
$34$&$1.60$&$0.44$\\
\hline
\multicolumn{3}{c}{{\bf Regen.}\hfill($L=4.4$~m, $\langle\lambda\rangle=500$~nm,  $N_{\rm pass}=200$)}\\[0.1cm]
\makebox[2.5cm][c]{$\theta\, [{\rm rad}]$}&\multicolumn{2}{c}{\makebox[5cm][c]{rate $[{\rm Hz}]$}}\\[0.1cm]
$0$&\multicolumn{2}{c}{$-0.012\pm0.009$}\\
$\frac{\pi}{2}$&\multicolumn{2}{c}{$0.013\pm0.007$}\\
\hline\hline
\end{tabular}
\end{center}

\vspace{3ex}
\begin{center}
\begin{tabular}{cc}
\hline\hline
\multicolumn{2}{c}{\makebox[8cm][c]{\bf { PVLAS} experiment}}\\
\hline
\multicolumn{2}{c}{{\bf Rotation}\hfill($L=1$~m, $N_{\rm pass}=44000$, $\theta=\frac{\pi}{4}$)}\\[0.2cm]
 \makebox[2.5cm][c]{$\lambda\,\, [{\rm nm}]$}&\makebox[5cm][c]{${\Delta\theta}\,[10^{-12}\,
{\rm rad}/{\rm pass}]$}\\ \hline\\[-0.45cm]
 $1064$&$(\pm ?) 3.9\pm0.2$\\
$532$&$+6.3\pm1.0$ {\bf (preliminary)}\\
\hline
\multicolumn{2}{c}{{\bf Ellipticity}\hfill($L=1$~m, $N_{\rm pass}=44000$, $\theta=\frac{\pi}{4}$)}\\[0.2cm]
\makebox[2.5cm][c]{$\lambda\,\, [{\rm nm}]$}&\makebox[5cm][c]{$\psi\,[10^{-12}\,{\rm rad}/{\rm pass}]$}
\\ \hline \\[-0.45cm]
$1064$&$-3.4\pm0.3$ {\bf (preliminary)}\\
$532$&$-6.0\pm0.6$ {\bf (preliminary)}\\
\hline\hline
\end{tabular}
\end{center}

\vspace{3ex}
\begin{center}
\begin{tabular}{ccc}
\hline\hline
\multicolumn{3}{c}{\makebox[8cm][c]{\bf { Q\&A} experiment}}\\
\hline
\multicolumn{3}{c}{{\bf Rotation}\hfill($L=1$~m, $\lambda=1064$~nm, $\theta=\frac{\pi}{4}$)}\\[0.1cm]
\makebox[2.5cm][c]{$N_{\rm pass}$}&\multicolumn{2}{c}{\makebox[5cm][c]{$\Delta\theta\,[{\rm nrad}]$}}\\[0.1cm]
$18700$&\multicolumn{2}{c}{$-0.4\pm5.3$}\\
 \hline\hline
\end{tabular}
\end{center}
\end{table}

\section{Vacuum Magnetic Dichroism and Birefringence}

The PVLAS collaboration is running a prime example for such an experiment at the INFN Legnaro
in Italy\cite{Zavattini:2005tm}. Similar experiments have been performed in the early nineties in Brookhaven 
(Brookhaven-Fermilab-Rochester-Trieste (BFRT) collabo\-ra\-tion\cite{Semertzidis:1990qc,Cameron:1993mr}) 
and are currently pursued also in Taiwan (Q\&A collaboration\cite{Chen:2006cd}) and in 
France (BMV collaboration\cite{Rizzo:Patras}).  In these experiments, 
linearly polarized laser photons are send through a superconducting dipole magnet 
(cf. Fig.~\ref{fig:pvlas}), with
the aim of measuring a change of the polarization state in the form of a possible 
rotation (vacuum magnetic dichroism)  and ellipticity (vacuum magnetic birefringence) 
(cf. Fig.~\ref{fig:pol_int}). Quite surprisingly and in contrast to the other experiments 
mentioned, PVLAS reported recently the observation of a 
quite sizeable vacuum magnetic dichroism\cite{Zavattini:2005tm} 
(cf. Table~\ref{tab:pol_data}). 
Moreover, preliminary data seem to indicate also evidence for an anomalously large 
vacuum magnetic birefringence (cf. Table~\ref{tab:pol_data}).  
These observations have led to a number of theoretical
and experimental activities, since the magnitude of the reported signals exceeds the
standard model expectations\cite{Adler:1971wn,Adler:2006zs,Biswas:2006cr} by far 
(see however Ref.~\cite{Davis:2007wu}).

\section{Possible Explanations}

Among possible particle physics 
explanations\cite{Maiani:1986md,Raffelt:1987im,Gies:2006ca,Kruglov:2007pg,Beswick:2007hq} 
of the reported signals two are particularly appealing in the sense that they can 
easily be embedded in popular extensions of the standard model:

The real and virtual production of 
\begin{itemize}
\item[(i)] a neutral spin-0 (axion-like) 
particle\cite{Maiani:1986md} (ALP) $\phi$
 with mass $m_\phi$ and a coupling to two photons via 
\begin{equation}
\label{scalar}
{\mathcal  L}^{(+)}_{\rm{int}}
  =-\frac{1}{4}g\phi^{(+)}F_{\mu\nu}F^{\mu\nu}
  =\frac{1}{2}g\phi^{(+)}(\vec{E}^{2}-\vec{B}^{2}),
\end{equation}
or
\begin{equation}
\label{pseudoscalar}
{\mathcal L}^{(-)}_{\rm{int}}
  =-\frac{1}{4}g\phi^{(-)}F_{\mu\nu}\widetilde{F}^{\mu\nu}
  =g\phi^{(-)}(\vec{E}\cdot\vec{B}),
\end{equation}
depending on its parity\footnote{For an analysis, where the ALP is not assumed to be 
an eigenstate of parity, see Ref.~\cite{Liao:2007nu}.}, 
denoted by the superscript $(\pm)$ (cf. Fig.~\ref{fig:pol_int} (top)), or
\item[(ii)] a pair of minicharged, $Q_\epsilon =\epsilon e$,
particles\cite{Gies:2006ca}  (MCP) $\epsilon^+ \epsilon^-$
with mass $m_\epsilon$, coupling to photons in the
usual way via the minimal substitution $\partial_\mu \to D_\mu \equiv \partial_\mu -{\rm i}\epsilon
e A_\mu$ in the Lagrangian (cf. Fig.~\ref{fig:pol_int} (bottom)).
\end{itemize}

\begin{figure}
\begin{center}
\psfig{file=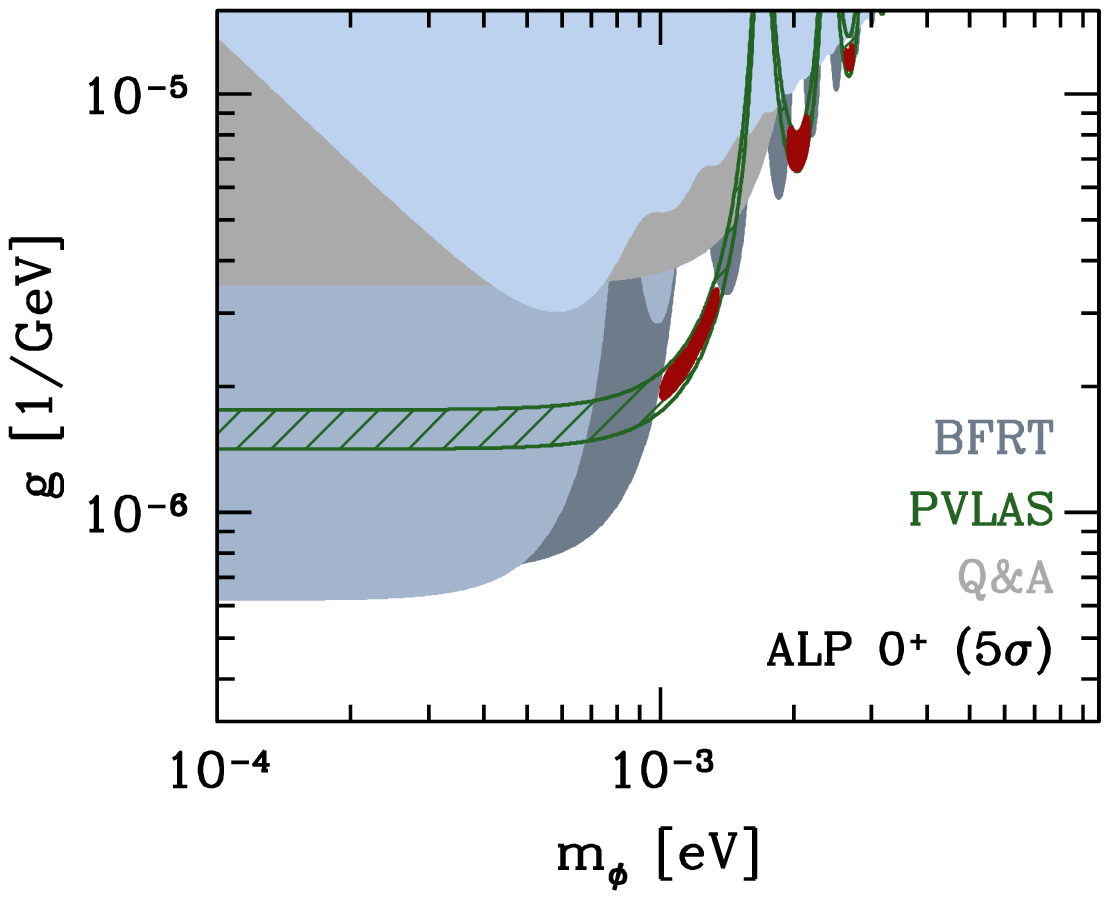,width=7.25cm}
\hfill
\psfig{file=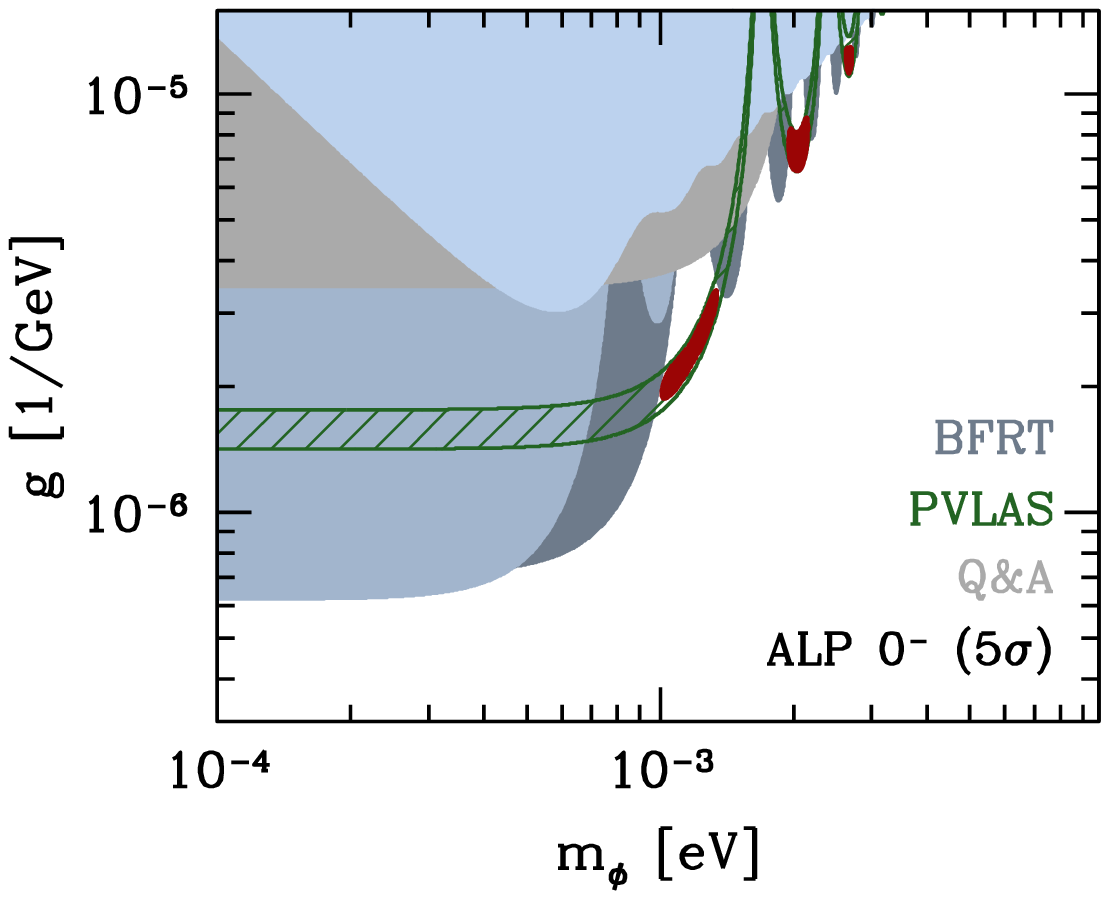,width=7.25cm}
\psfig{file=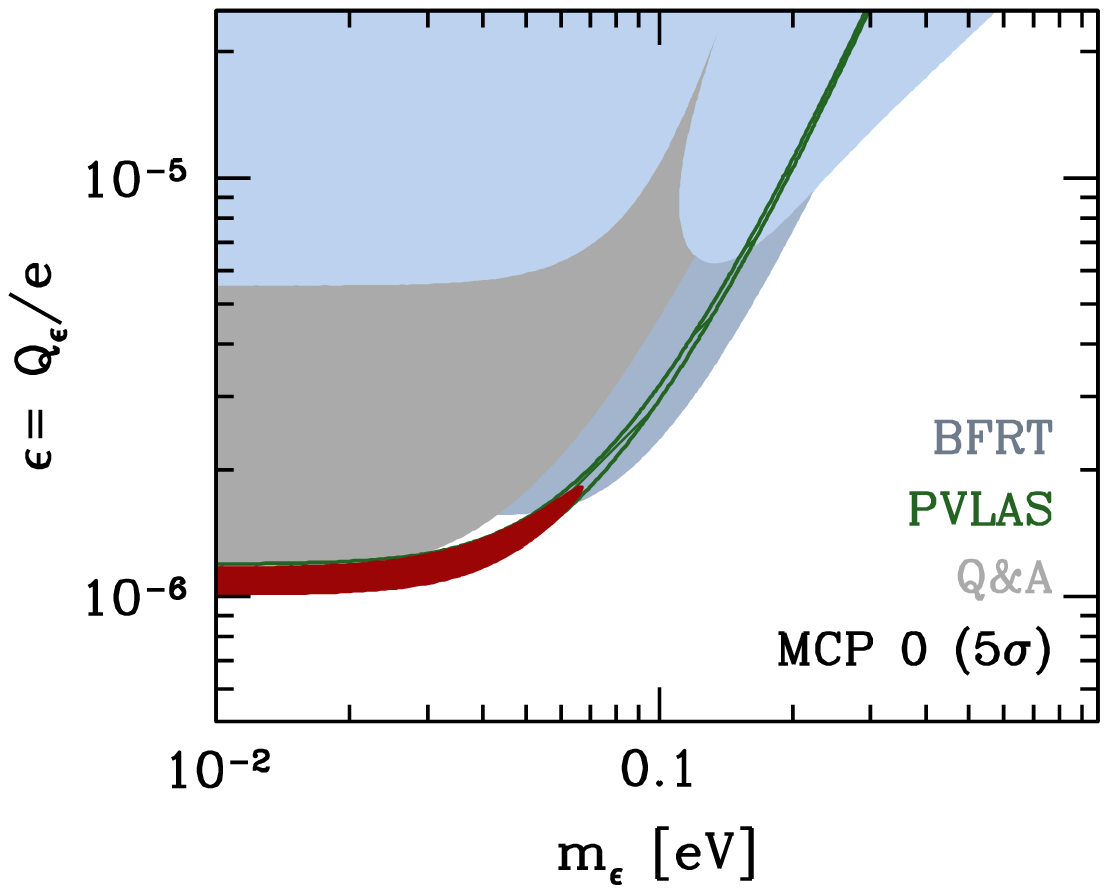,width=7.25cm}
\hfill
\psfig{file=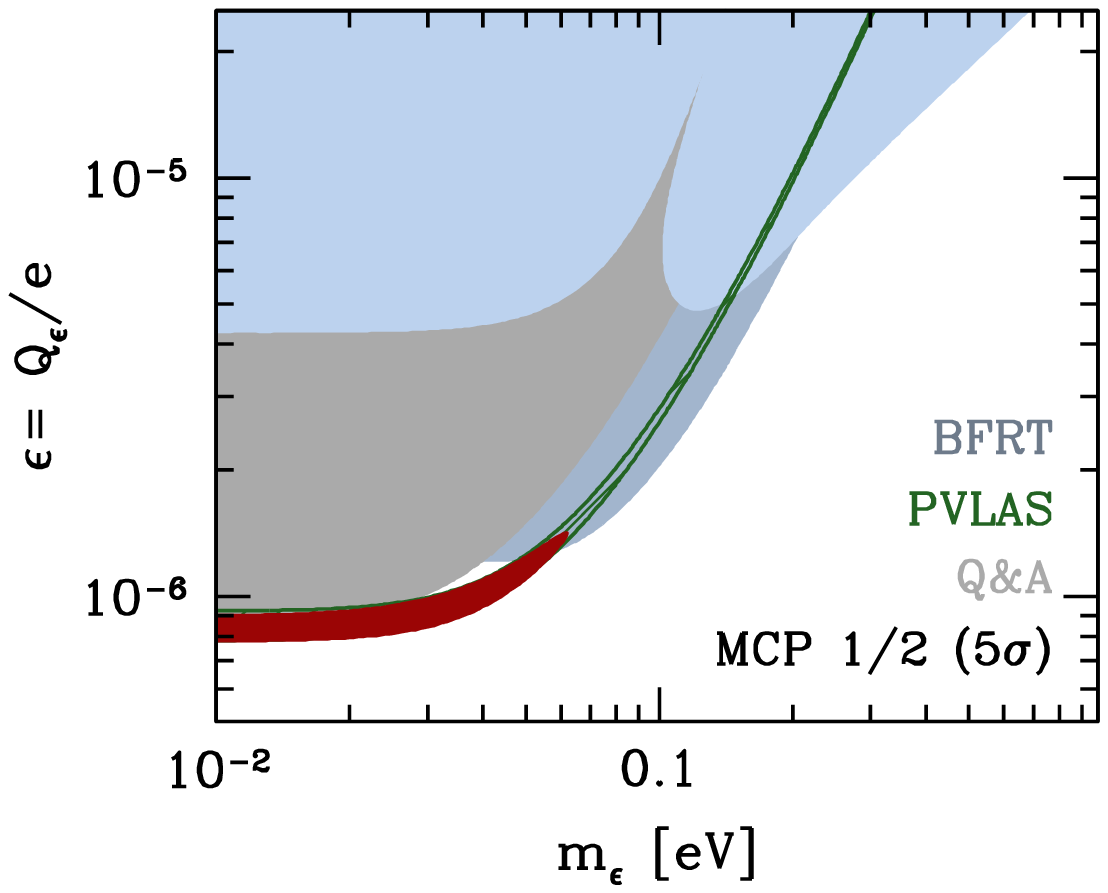,width=7.25cm}
\psfig{file=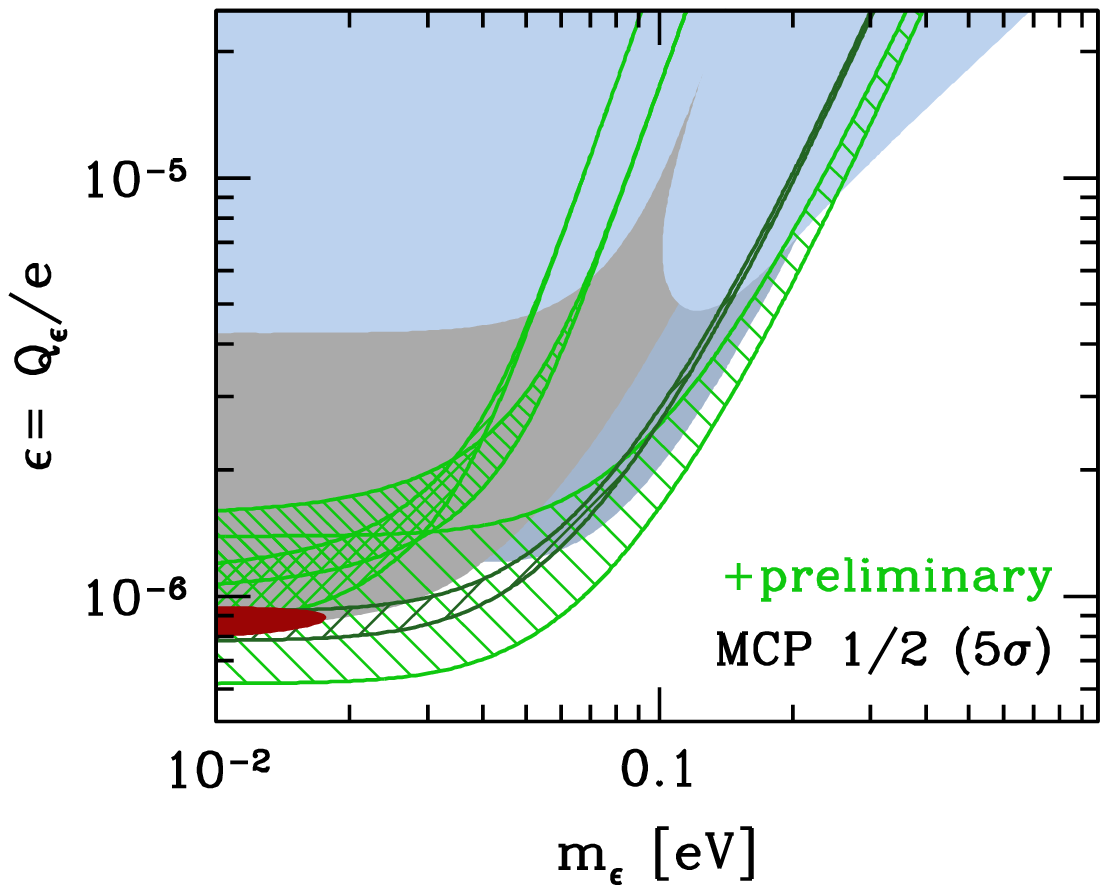,width=7.25cm}
\end{center}
\caption[]{Pure ALP scalar (pseudo-scalar) (top left (right)) and 
  pure MCP spin-$0$ ($1/2$) (middle left (right) and bottom) interpretation of the data on vacuum magnetic 
  dichroism,
  birefringence and photon regeneration\cite{Ahlers:2006iz}: 
  $5\sigma$ confidence level of the model parameters
  (red). The blue-shaded regions arise from the BFRT upper limits\cite{Cameron:1993mr} for
  regeneration (dark blue), rotation (blue) and ellipticity (light
  blue).  The gray-shaded region is the Q\&A upper limit\cite{Chen:2006cd} for rotation.
  The dark-green band shows the published result of PVLAS for rotation\cite{Zavattini:2005tm} with
  $\lambda=1064$~nm. The bottom panel includes also the   
  $5\sigma$ C.L.s for rotation (coarse hatched) and ellipticity (fine
  hatched) with $\lambda=532$~nm (left hatched) and $\lambda=1064$~nm
  (right hatched), respectively, from the preliminary PVLAS data (cf. Table~\ref{tab:pol_data}).} 
\label{fig:part_int}
\end{figure}

Indeed, as apparent from Fig.~\ref{fig:part_int} (top), the 
rotation observed by PVLAS can be reconciled with the non-observation
of a signal by BFRT and Q\&A, if there is an ALP 
with\cite{Zavattini:2005tm} a mass $m_\phi\sim$~meV and a coupling
$g\sim 10^{-6}$~GeV$^{-1}$. 
Alternatively, the currently published experimental data are compatible with the 
existence of an MCP with\cite{Gies:2006ca} $m_\epsilon\lwig 0.1$~eV and $\epsilon \sim 10^{-6}$ 
(cf. Fig.~\ref{fig:part_int} (middle)). 

This degeneracy can be lifted eventually by including more data from different 
experimental settings from the PVLAS collaboration. As an illustration, one may include 
the preliminary PVLAS data from Table~\ref{tab:pol_data}. It is easily seen that the signs of the 
rotation and the ellipticity are incompatible with a pure scalar ($0^+$) ALP, a pure 
pseudo-scalar ($0^-$) ALP, and a pure MCP spin-$0$ interpretation\cite{Ahlers:2006iz}.  
They prefer a pure MCP spin-$1/2$ interpretation (cf. Fig.~\ref{fig:part_int} (bottom)). 
A slightly better fit is found\cite{Ahlers:priv} from a 
combination of ALP $0^+$ plus MCP $1/2$.

\section{Crucial Laboratory Tests}

It is very comforting that a number of laboratory-based\footnote{For astrophysics-based 
tests of the ALP interpretation of the PVLAS anomaly see 
Refs.~\cite{Dupays:2005xs,Fairbairn:2006hv,Mirizzi:2007hr}.} low-energy\footnote{High-energy collider-based
tests do not seem to be competitive in the near future\cite{Kleban:2005rj}.}
tests of the ALP and MCP interpretation of the PVLAS anomaly are
currently set up and expected to yield decisive results within the
upcoming year. For example, in addition to PVLAS, the  Q\&A\cite{Chen:2006cd},  
BMV\cite{Rizzo:Patras}, and later the OSQAR\cite{Pugnat:2005nk,OSQAR} collaborations 
will run further polarization experiments with different experimental parameter 
values which finally may lead to a  discrimination between the ALP and the MCP 
hypothesis\cite{Ahlers:2006iz}. 

\begin{figure}
\begin{center}
\psfig{file=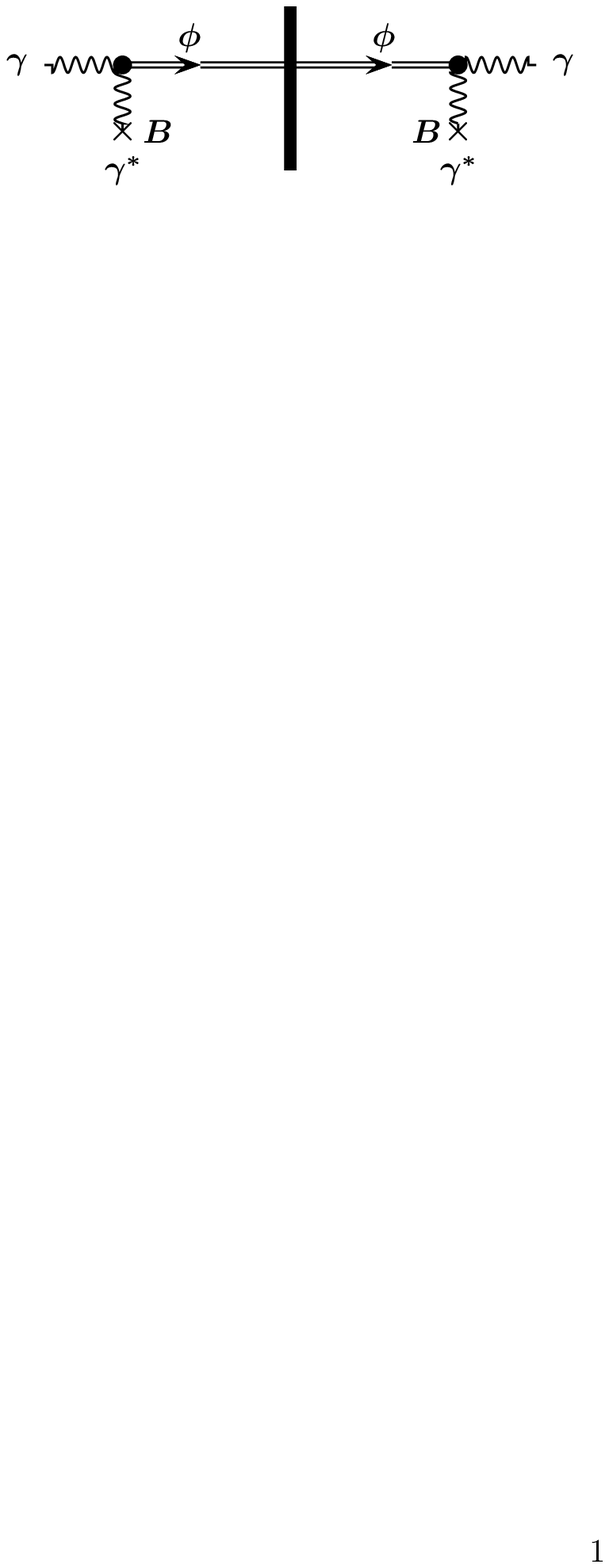,bbllx=86pt,bblly=637pt,bburx=298pt,bbury=707pt,width=11cm}

\psfig{file=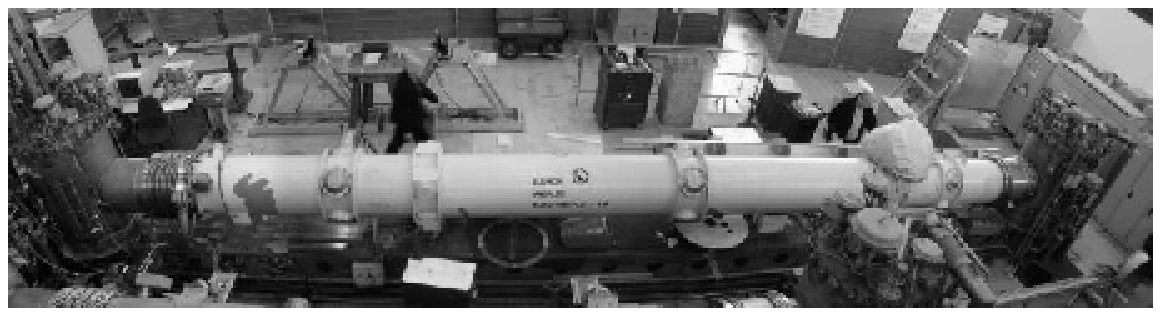,width=14cm}
\end{center}
\caption[]{Light shining through a wall. 
{\em Top:} Schematic view of ALP production through photon conversion 
in a magnetic field (left), subsequent travel through a wall, and final 
detection through photon regeneration (right).
{\em Bottom:} Superconducting HERA dipole magnet exploited for light shining through a wall 
in the Axion-Like Particle Search (ALPS) 
experiment\cite{Ehret:2007cm}, a collaboration between DESY, Laser Zentrum 
Hannover and Sternwarte Bergedorf.}
\label{fig:reg}
\end{figure}

\subsection{Light Shining Through a Wall}

The ALP interpretation of the PVLAS signal will crucially be tested by photon
regeneration (sometimes called ``light shining through walls'')
experiments\cite{Sikivie:1983ip,Anselm:1986gz,VanBibber:1987rq,Ringwald:2003ns,Rabadan:2005dm,Sikivie:2007qm}, 
presently 
under construction or serious 
consideration\cite{Rizzo:Patras,OSQAR,Ehret:2007cm,Baker:Patras,Cantatore:Patras} 
(cf. Table~\ref{tab:exp}).  
In these experiments (cf. Fig.~\ref{fig:reg}), a photon beam is directed across
a magnetic field, where a fraction of them turns into ALPs. 
The ALP
beam can then propagate freely through a wall or another obstruction
without being absorbed, and finally another magnetic field located on
the other side of the wall can transform some of these ALPs into
photons --- seemingly regenerating these photons out of nothing.
A pioneering photon regeneration experiment has been done also by the
BFRT collaboration\cite{Cameron:1993mr,Ruoso:1992nx}. 
No signal has been found and the corresponding upper limit on 
$g$ vs. $m_\phi$ is included in Fig.~\ref{fig:part_int} (top). 
In Hamburg, the Axion-Like Particle Search (ALPS) collaboration between 
DESY, Laser Zentrum Hannover and Sternwarte Bergedorf is presently setting
up such an experiment (cf. Fig.~\ref{fig:reg} (bottom)) which will
take data in summer 2007 and firmly establish or exclude the
ALP interpretation of the PVLAS data.
  
As an incidental remark let us note an obvious, but remarkable spin-off 
if a positive signal  is detected in one of the light shining through a wall experiments mentioned  above. 
It would provide the proof of principle of an ``ALP beam radio'' -- based on the possibility to 
send signals through material which is untransparent to photons --  
as a means of long-distance, possibly world-wide telecommunication. 
With presently available technology, however, only
a very low signal transmission rate may be achieved\cite{Stancil:2007yk}.

\begin{table}
\caption[]{Experimental parameters of upcoming photon regeneration experiments: 
magnetic fields $B_i$ and their length $\ell_i$ on production ($i=1$) and 
regeneration ($i=2$) side (cf. Fig.~\ref{fig:reg}); 
and the corresponding photon conversion and reconversion probability $P_{\gamma\phi\gamma}$, 
for $g\sim 2\times 10^{-6}$~GeV$^{-1}$.}
\begin{center}
\begin{tabular}{|l|l|c||l||}
\hline
Name & Laboratory   & Magnets & $P_{\gamma\phi\gamma|g\sim 2\times 10^{-6}/{\rm GeV}}$  \\
\hline
{\bf ALPS}\cite{Ehret:2007cm}&DESY/D     & $B_1=B_2=5$~T & 
\\
 &  	      & $\ell_1=\ell_2=4.21$~m &  $\sim 10^{-19}$  \\
\hline
{\bf BMV}\cite{Rizzo:Patras} &LULI/F 	    & $B_1=B_2=11$~T &
\\
 &        & $\ell_1=\ell_2=0.25$~m &  $\sim 10^{-21}$  \\
\hline 
{\bf LIPSS}\cite{Baker:Patras}&Jlab/USA    & $B_1=B_2=1.7$~T & 
\\
 &    & $\ell_1=\ell_2=1$~m & $\sim 10^{-23.5}$  \\
\hline
{\bf OSQAR}\cite{OSQAR}&CERN/CH    & $B_1=B_2=11$~T & 
\\
 &        & $\ell_1=\ell_2=7$~m &  $\sim 10^{-17}$ \\
\hline
 &    & $B_1=5$~T & 
\\
{\bf PVLAS}\cite{Cantatore:Patras} & Legnaro/I     & $\ell_1=1$~m &   $\sim 10^{-23}$
\\
	 &  			& $B_2=2.2$~T & \\
&            &$\ell_2=0.5$~m  &\\
\hline
\end{tabular}
\end{center}
\label{tab:exp}
\end{table}

\subsection{Dark Current Flowing Through a Wall}

Clearly, light shining through a wall in the above set up will be negligible in pure MCP models,  
since the probability that the $\epsilon^\pm$ pairs produced before the wall meet again and
recombine behind the wall will be negligible. 
However, one may exploit in this case Schwinger pair-production of MCPs 
in the strong electric fields available in accelerator cavities\cite{Gies:2006hv}.
This will lead to a new form of energy loss. In fact, one of the best 
current laboratory limits on very light MCPs, $\epsilon < 10^{-6}$ for 
$m_\epsilon \lwig 0.1$~meV,  arises from the fact that the superconducting cavities 
of the type developed for the Tera Electronvolt Superconducting Linear Accelerator (TESLA) 
have a very high quality factor\cite{Lilje:2004ib}, corresponding to 
a very low energy loss. 
A more direct approach to infer the existence of such particles may be based on the detection
of the macroscopic electrical current comprised of them in the form of a 
``dark current flowing through a wall'' experiment\cite{Gies:2006hv}. 
\begin{figure}[ht]
\begin{center}
\psfig{file=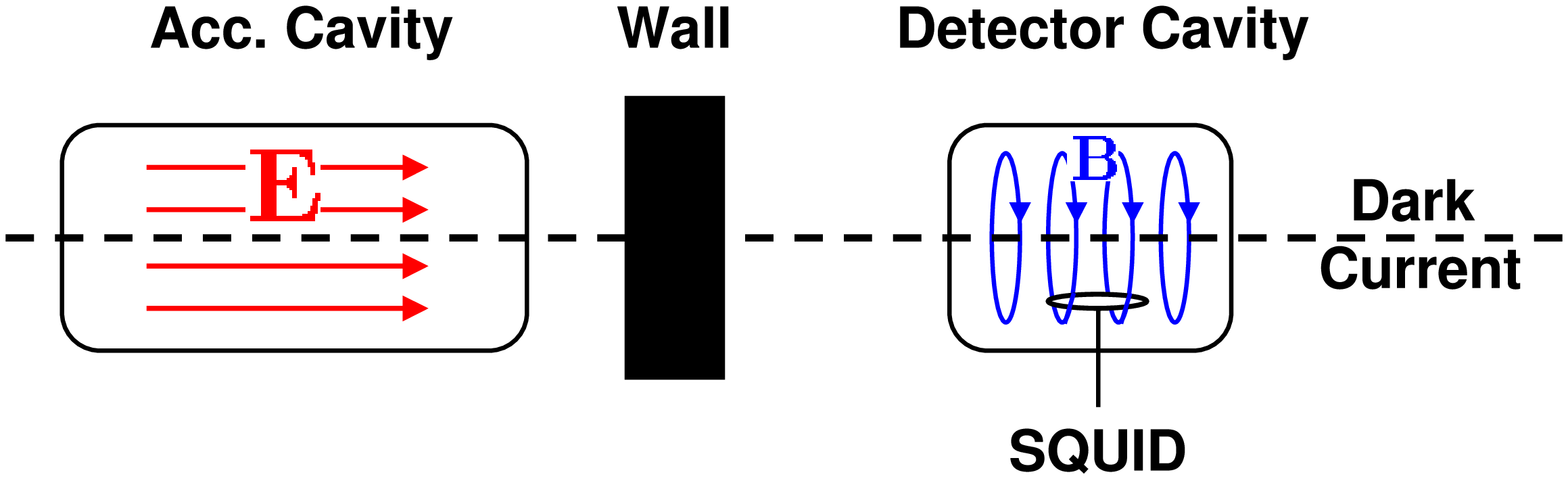,bbllx=136,bblly=71,bburx=339,bbury=738,angle=-90,width=11.5cm,clip=}

\vspace{2ex}
\psfig{file=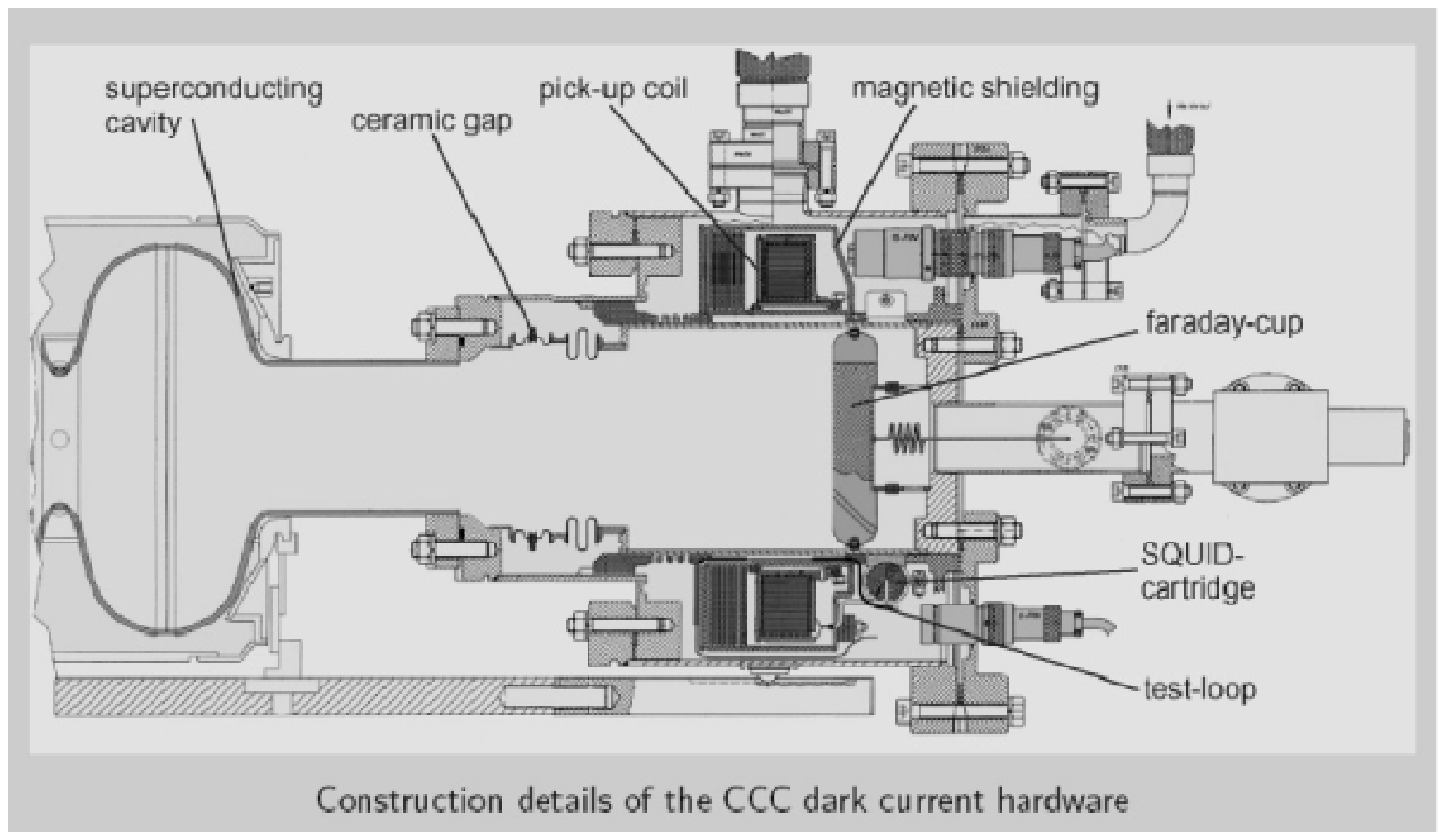,width=11.5cm,clip=}
\end{center}
\caption[...]{Dark current flowing through a wall. 
{\em Top:} Schematic set up for a ``dark current flowing through a wall'' 
experiment. The alternating dark current (frequency $\nu$), comprised of the produced millicharged particles 
(dashed line), escapes from the accelerator cavity  and traverses also a 
thick shielding (``wall''), in which the conventional dark current of electrons
is stopped. The dark current induces a magnetic field in a resonant (frequency $\nu$) detector cavity  
behind the wall, which is detected by a SQUID\cite{Gies:2006hv}. 
{\em Bottom:} Proposed set up for an absolute measurement of the dark current 
from a TESLA superconducting accelerator cavity with the help of a 
cryogenic current comparator\cite{Vodel:2005ma,Wendt:CCC}.     
\label{fig:detector}}
\end{figure}
In Fig.~\ref{fig:detector} (top), 
we show schematically how one could set up an experiment to detect this current. 
In fact, a collaboration between DESY, GSI, and the University of Jena has 
already developed\cite{Vodel:2005ma} 
a so-called cryogenic current comparator (CCC) (cf. Fig.~\ref{fig:detector} (bottom))
for the absolute measurement of the dark currents leaving the TESLA cavities down to values 
of pA. Placing an absorber between the TESLA cavity in 
Fig.~\ref{fig:detector} (bottom) and the CCC, one may realize easily a 
dark current flowing through a wall experiment. An exclusion of a 
dark current of size $\mu$A (nA) will result in a limit\cite{Jaeckel:priv}  
$\epsilon < 10^{-6}$ ($10^{-7}$) for very light MCPs, $m_\epsilon < 0.1$~meV.     

We note in passing that the eventual experimental demonstration that a dark current, produced 
in an accelerator cavity, flows through a wall and can be detected
behind the wall would indicate the exciting possibility of an ``MCP beam radio'' 
as a new-type of telecommunication,  in analogy to the ALP beam radio mentioned above,

\subsection{Search for Invisible Orthopositronium Decay}

A classical probe for MCPs is the search
for invisible orthopositronium (OP) decays\cite{Dobroliubov:1989mr,Mitsui:1993ha}. Recently, the ETH-INR 
collaboration published\cite{Badertscher:2006fm} a new stringent limit on the branching ratio 
${\rm Br}({\rm OP}\to {\rm invisible})<4.2\times 10^{-7}$, 
which translates, on account of the prediction 
${\rm Br}({\rm OP}\to \epsilon^+\epsilon^-)\simeq 371\,\epsilon^2$, for $m_\epsilon\ll m_e$,
into a limit $\epsilon<3.4\times 10^{-5}$ on the fractional charge of the MCPs $\epsilon^\pm$. 
Further improvements and other experiments are 
being developed\cite{Rubbia:2004ix,Vetter:2004fs}, 
which may reach finally a sensitivity of $10^{-10}$ in the 
branching ratio ${\rm Br}({\rm OP}\to {\rm invisible})$, corresponding to 
a sensitivity of $5\times 10^{-7}$ in $\epsilon$, seriously probing the 
MCP interpretation of the PVLAS data\footnote{The search for the Lamb shift 
contribution of light MCPs does not seem to be competitive with the 
search for invisible OP decays: it yields a weaker limit\cite{Gluck:2007ia}, $\epsilon < 10^{-4}$,  
for $m_\epsilon\,\lwig\,1$\,keV.}. 

\subsection{Searches Near Nuclear Reactors}

Another method to infer the existence of MCPs is the search for
excess electrons from elastic $\epsilon^\pm$ scattering in a detector
near a nuclear reactor. Indeed, nuclear reactors with power exceeding
$2$~GW emit more than $10^{20}$ photons per second, which may 
partially convert into $\epsilon^\pm$ pairs within the reactor core. 
A small fraction of these particles could lead to an observable excess
of electrons via the above mentioned elastic scattering process. 
Recent corresponding results from the TEXONO experiment set up at the 
Kuo-Sheng Nuclear Power Station ($2.8$~GW), originally given in terms of 
bounds on the magnetic dipole moment of neutrinos, can be translated into 
a bound\cite{Gninenko:2006fi} $\epsilon <10^{-5}$, for $m_\epsilon\,\lwig\,$keV,
which is only about one order of magnitude below the required sensitivity
to test the pure MCP interpretation of PVLAS.  
This bound may be improved in the near future by exploiting a massive
liquid Argon detector.    

\begin{figure}
\begin{center}
\psfig{file=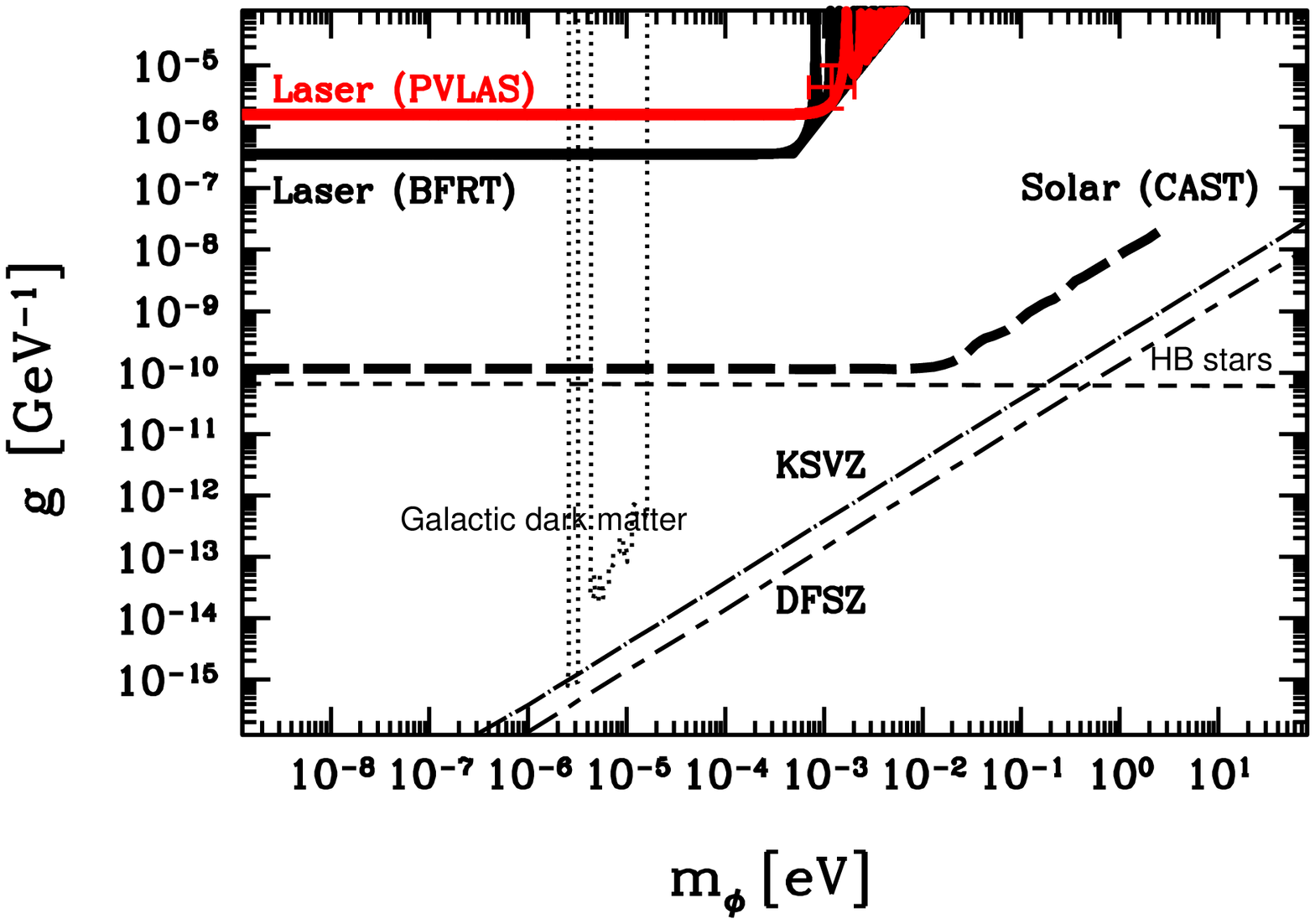,bbllx=25,bblly=224,bburx=564,bbury=603,width=10.5cm,clip=}

\vspace{2ex}
\psfig{file=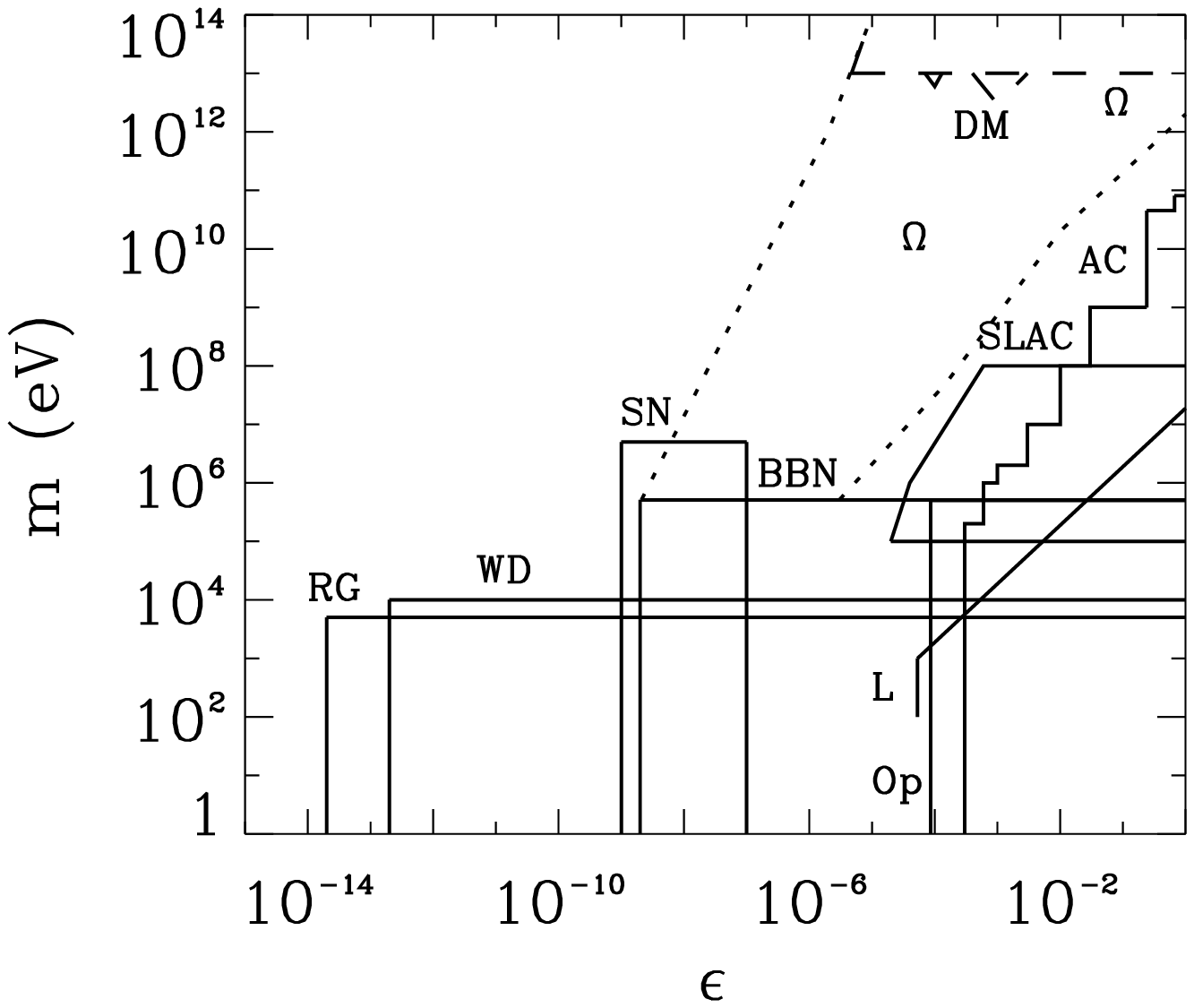,bbllx=102,bblly=380,bburx=477,bbury=702,width=9.cm,clip=}
\end{center}
\caption[...]{Constraints on ALP (top) and MCP (bottom) parameters. 
{\em Top:} Upper limits on ALP coupling $g$ vs. its mass $m_\phi$. 
The laser experiments\cite{Zavattini:2005tm,Cameron:1993mr} aim at  
$\phi$ production and detection in the laboratory.  
The galactic dark matter experiments exploit microwave cavities to detect 
ALPs  under the assumption that they are the dominant constituents of our 
galactic halo\cite{Yao:2006px}, and the solar experiments 
search for ALPs from the sun\cite{Zioutas:2004hi}. 
The constraint from horizontal branch (HB) stars\cite{Raffelt:1996}  
arises from a consideration of stellar  
energy losses through ALP production.
The predictions from two quite distinct QCD axion models, namely the KSVZ\cite{Kim:1979if,Shifman:1980if} 
(or hadronic) and the DFSZ\cite{Zhitnitsky:1980tq,Dine:1983ah} (or grand unified) one, 
are also shown.      
{\em Bottom:} Exclusion regions in MCP fractional electric charge $\epsilon$ vs. mass $m=m_\epsilon$
(from Ref.~\cite{Davidson:2000hf}). 
The bounds arise from the following constraints: AC -- accelerator experiments; Op -- the Tokyo 
search for the invisible decay of orthopositronium\cite{Mitsui:1993ha}; 
SLAC -- the SLAC minicharged particle  search\cite{Prinz:1998ua}; 
L -- Lamb shift; BBN -- nucleosynthesis; $\Omega$ -- $\Omega < 1$; 
RG -- plasmon decay in red giants; WD -- plasmon decay in white dwarfs; 
DM -- dark matter searches; SN -- supernova 1987A.
\label{fig:astro_constr}}
\end{figure}

\section{Problems of Particle Interpretations}

\subsection{Constraints from Astrophysics and Cosmology} 

Both, the ALP as well as the MCP interpretation of the PVLAS data 
seem to be in serious conflict with astrophysical bounds, arising 
from energy loss considerations of stars\cite{Raffelt:1996,Davidson:2000hf}. 

ALP production due to Primakoff processes $\gamma Z\to \phi Z$ in the stellar plasma 
and subsequent ALP escape would lead to drastic changes in the timescales
of stellar evolution, placing a bound $g<8\times 10^{-11}$~GeV$^{-1}$ for 
$m_\phi\,\lwig\,$keV, slightly stronger than the published bound arising from the non-observation of
photon conversion of ALPs, eventually produced in the sun, by the CERN Axion Solar 
Telescope CAST\cite{Zioutas:2004hi} (cf. Fig.~\ref{fig:astro_constr} (top)). 
These bounds on $g$ are more than four orders of magnitude smaller than the 
values suggested by a pure ALP interpretation of PVLAS. 
This serious conflict may be solved if the production of ALPs 
is heavily suppressed\footnote{For alternative proposals to solve this conflict based on trapping of ALPs
within stellar cores  
see Refs.\cite{Jain:2005nh,Jain:2006ki}.} 
 in astrophysical 
plasmas\cite{Masso:2005ym,Jaeckel:2006id,Jaeckel:2006xm,Brax:2007ak}, i.e. if
$g_{\rm |plasma}\ll g_{\rm vacuum}$.  
Interestingly enough, microphysical models achieving such a 
suppression require typically even more sub-eV 
particles and fields\cite{Masso:2006gc,Mohapatra:2006pv}. 
 
In the case of MCPs, a prominent production mechanism in stellar plasmas
is plasmon decay, $\gamma^\ast \to \epsilon^+ \epsilon^-$, which is effective
as soon as the plasma frequency $\omega_{\rm p}\sim $~few keV exceeds the threshold
for pair production, $2m_\epsilon$. The lifetime of red giants leads to 
the most stringent bound $\epsilon < 2\times 10^{-14}$, for $m_\epsilon\,\lwig\,5$\,keV,
on the fractional electric charge, considerably stronger than the
bound arising from big bang nucleosynthesis (cf. Fig.~\ref{fig:astro_constr} (bottom)).
The red giant bound on $\epsilon$ is thus more than seven orders of magnitude below the
value required by a pure MCP interpretation of PVLAS. Again, a reconciliation
can be achieved if the effective charge in the plasma is much smaller than 
in vacuum, i.e. $\epsilon_{\rm |plasma}\ll \epsilon_{\rm vacuum}$ -- as for example in the 
models\cite{Masso:2006gc,Abel:2006qt,Foot:2007cq} discussed in the next section.  

Recently, it has been pointed out that the production of sub-eV mass MCPs through 
collisions of cosmic microwave background (CMB) photons, $\gamma +\gamma \to \epsilon^+ + \epsilon^-$, 
may distort the CMB energy spectrum\cite{Melchiorri:2007sq}. From a comparison with
the observed spectrum, a limit $\epsilon<10^{-7}$ is inferred. This is about one order
of magnitude below the value required in a pure MCP interpretation of PVLAS 
(cf. Fig.~\ref{fig:part_int} (middle and bottom)). It remains to be seen whether this 
apparent conflict can be reconciled in certain models.

\begin{figure}
\begin{center}
\psfig{file=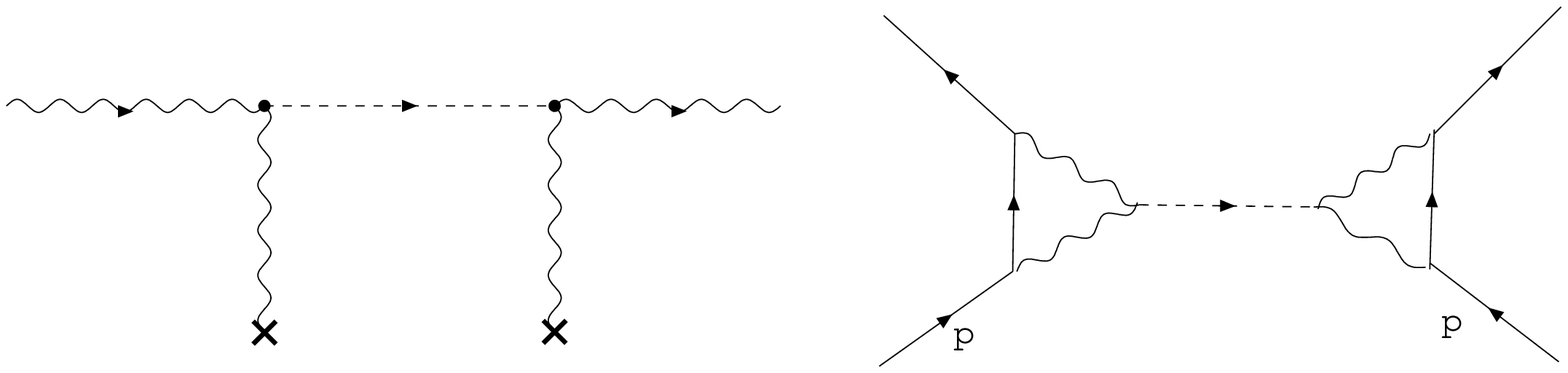,bbllx=337,bblly=0,bburx=626,bbury=149,width=10.5cm,clip=}
\end{center}
\caption[...]{Exchange of a scalar ALP, coupled two photons via Eq.~(\ref{scalar}), 
between two protons, giving rise to a Yukawa-type non-Newtonian force between two neutral test bodies 
(from Ref.~\cite{Adelberger:2006dh}).  
\label{fig:non_newtonian}}
\end{figure}

\subsection{Constraints from Searches for Non-Newtonian Forces}

A scalar ALP will couple radiatively to protons, 
leading to a spin-independent non-Newtonian
force between test bodies of the Yukawa-type, $\propto (gm_p)^2\exp(-m_\phi r)$ 
 (cf. Fig.~\ref{fig:non_newtonian}). 
From the non-observation of such a force in sensitive torsion-balance
searches for Yukawa violations of the gravitational inverse-square law 
one may put a very stringent limit\cite{Dupays:2006dp,Adelberger:2006dh}, 
$g<4\times 10^{-17}$~GeV$^{-1}$, for $m_\phi = 1$~eV and assuming that the effective 
interaction Eq.~(\ref{scalar}) is valid up high energies $\Lambda\gg m_p\sim 1$~GeV.
This limit seemingly rules out completely the 
ALP $0^+$ interpretation of the PVLAS data, which requires a coupling in the 
$10^{-6}$~GeV$^{-1}$ range. However, in models where new 
physics arises already at sub-eV scales, $\Lambda\sim$~meV, this 
strong conflict may relax very much, as we will see in the next section.

\section{\label{sec:model_building}WILPs in Models with Light Extra-U(1)'s}

Finally, let us consider in this section a class of models 
\begin{itemize}
\item in which MCPs with $\epsilon\ll 1$ arise naturally, 
\item which may be easily embedded in popular extensions of the standard model, 
\item in which most of the conflicts with 
astrophysics, cosmology, etc. can be evaded.
\end{itemize} 

Particles with small, unquantized charge arise very naturally in so-called
paraphoton models\cite{Okun:1982xi}, containing, beyond the usual ``visible'' electromagnetic U(1)
gauge factor additional ``hidden'' U(1) factors, which may kinetically mix with the 
visible one. Such hidden-sector U(1)'s and their mixing occur in 
many extensions of the standard model, in particular in those coming from string theory. 
The crucial observation is that particles charged under the hidden U(1)'s 
get an induced visible electric charge proportional to the kinetic mixing parameter\cite{Holdom:1985ag}.

As a specific enlightening example\cite{Masso:2006gc}, let us consider a gauge theory model with three 
light Abelian gauge fields $A_i$, $i=0,1,2$, described by three U(1) factors, 
${\rm U}_0(1)\times {\rm U}_1(1)\times {\rm U}_2(1)$, which interact with charged matter fields, 
entering the currents $j_i$, $i=0,1,2$. Exploiting a matrix notation for the gauge fields, 
$A\equiv (A_0, A_1, A_2)^T$, and their field strength, $F \equiv (F_0, F_1, F_2)^T$,
the Lagrangian, in the basis where the interactions with charged fields is diagonal, 
can be written as
\begin{equation}
{\cal L} =-\frac{1}{4}\, F^T {\cal K}_F\,  F+ \frac{1}{2}\, A^T
{\cal M}_A^2\,  A + e\sum_i\, j_{i} A_i 
\label{L_complete}
\,. 
\end{equation}
Here, $j_0$ is assumed to be constructed from the fields corresponding to our visible charged 
standard model  
particles, whereas $j_1$ and $j_2$ are assumed to be constructed from the fields
corresponding to the hidden-sector exotic particles. 
We assume that there are small mixing terms in the gauge kinetic matrix ${\cal K}_F$ and that
the masses of the paraphotons entering the mass matrix ${\cal M}_A^2$ are small. Specifically, 
\begin{equation}
 {\cal K}_F  = \left( \begin{array}{ccc}
                                            1 & \chi & \chi  \\
                                            \chi & 1 &   0          \\
                                            \chi & 0 & 1
 \end{array}\right)
\,,\hspace{6ex}
 {\cal M}_A^2  = \left( \begin{array}{ccc}
                                            0 & 0 & 0  \\
                                            0 & \mu^2 &   0          \\
                                            0 & 0 & 0
 \end{array}\right)
\,,
 \label{MF}
\end{equation}
with small mixing parameters, $\chi\ll 1$.  
%
\begin{figure}
\begin{center}
\psfig{file=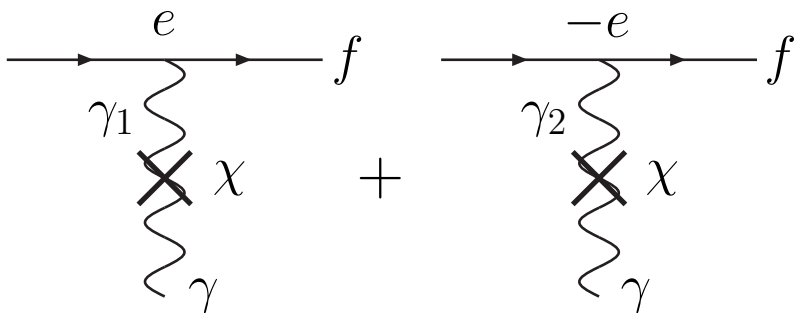,width=10.5cm,clip=}
\end{center}
\caption[...]{Gauge-kinetic mixing induced coupling of a hidden-sector 
              particle $f$, with charge assignments $(0,e,-e)$ under the gauge group  
              ${\rm U}_0(1)\times {\rm U}_1(1)\times {\rm U}_2(1)$, to a photon\cite{Ahlers:priv}. 
\label{fig:charge_shift}}
\end{figure}
%
From here, it is easily seen that a hidden-sector charged particle will experience 
a tiny visible-sector electric charge. Indeed, the effective coupling of  
a hidden-sector particle $f$ with charge assignment $(0,e,-e)$ to a visible-sector
photon with four-momentum squared $q^2$ can be easily read-off from Fig.~\ref{fig:charge_shift},
leading to an effective fractional electric charge\cite{Masso:2006gc}  
\begin{eqnarray}
\epsilon_f  \simeq \frac{\mu^2}{q^2-\mu^2}\, \chi \simeq 
\left\{
\begin{array}{rll}
- \chi & {\rm for} & q^2 = 0 \\
(\mu^2/q^2)\,\chi & {\rm for} & q^2 \gg \mu^2
\end{array}
\right.
\,,
\label{eff_charge}
\end{eqnarray}
which is naturally small, as long as the gauge kinetic mixing parameter $\chi\ll 1$. 

\begin{figure}
\begin{center}
\psfig{file=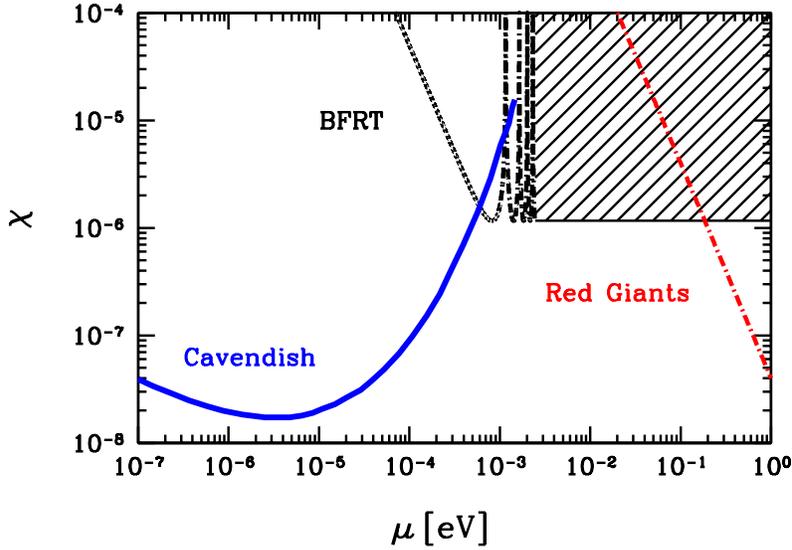,bbllx=32,bblly=227,bburx=578,bbury=608,width=10.5cm,clip=}
\end{center}
\caption[...]{Upper limit on the mixing parameter $\chi$ as a function of the mass $\mu$ of 
a light hidden-sector paraphoton. The limits arise from: Cavendish -- searches for deviations
from Coulomb's law\cite{Williams:1971ms,Bartlett:1988yy}; 
BFRT -- light shining through a wall\cite{Cameron:1993mr} (in vacuum, without magnetic field); 
Red Giant -- plasmon decay $\gamma^\ast \to f\bar f$. 
\label{fig:paraph_lim}}
\end{figure}

This model, for $\chi\sim 10^{-6}$, therefore, readily reproduces the MCP interpretation of PVLAS.
Moreover, the conflict with the lifetime of stars can be relaxed by chosing $\mu$ in the sub-eV 
range\cite{Abel:2006qt}, $\mu\,\lwig\,0.1$\,eV. In fact, in the stellar plasma, the four-momentum squared 
$q^2=\omega_{\rm p}^2\sim {\rm keV}^2$ of the plasmon $\gamma^\ast$ is in this case large enough that 
the additional suppression factor $\mu^2/\omega_{\rm p}^2\,\lwig\,10^{-8}$ 
in Eq.~(\ref{eff_charge}) leads to
a reconciliation of the PVLAS suggested value for $|\epsilon_f(q^2=0)|\simeq \chi\sim 10^{-6}$ with the
requirement that in stellar plasma $\epsilon_f\,\lwig\,10^{-14}$. 
Further constraints on the paraphoton parameters of such a model can be obtained from 
Cavendish-type searches for deviations from Coulomb's law and from searches for 
light-shining through a wall, exploiting vacuum oscillations of photons into hidden-sector
paraphotons\cite{Okun:1982xi}. As apparent from Fig.~\ref{fig:paraph_lim}, the pioneering 
experiments of this type have already nearly reached the sensitivity to 
probe for the required paraphotons. Values of $\chi\,\lwig\,10^{-6}$ in the meV -- 0.1 eV mass range
may readily be probed by the next-generation of light shining through a wall experiments, which, 
in the case of photon-paraphoton oscillations and in contrast to the case of photon-ALP oscillations, 
require only high initial photon fluxes, but no external magnetic field, since they occur, for 
finite paraphoton mass, already in vacuum.  
Therefore,  should light paraphotons exist, the corresponding ``paraphoton beam radio''  
seems to offer the cheapest way of WILP-based telecommunication.

\begin{figure}
\begin{center}
\psfig{file=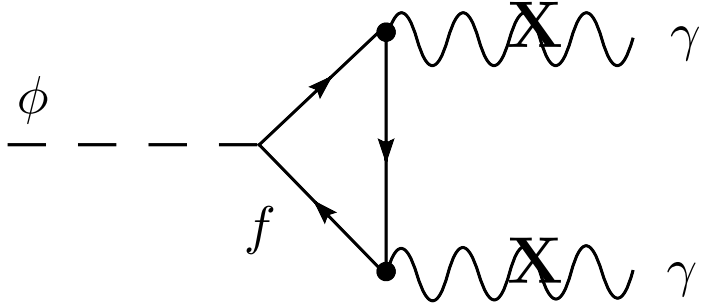,width=10.5cm,clip=}

\vspace{-4.8cm}
\hspace{4.5cm}
{\Huge\bf X}

\vspace{2.5cm}
\hspace{4.5cm}
{\Huge\bf X}

\end{center}
\caption[...]{Effective coupling of a hidden-sector spin-$0$ boson $\phi$ to two 
photons via a loop of hidden-sector paracharged fermions $f$ in a model with
gauge kinetic mixing\cite{Masso:2006gc}. 
\label{fig:triangle}}
\end{figure}

This class of minimal\cite{Abel:2006qt} models 
for explaining PVLAS may be extended\cite{Masso:2006gc} by introducing 
a light hidden sector spin-$0$ boson $\phi$, with a Yukawa coupling $y_f$ to the 
hidden-sector paracharged particle $f$. The corresponding radiatively induced 
coupling to two photons (cf. Fig.~\ref{fig:triangle}) can be arranged to be in 
the PVLAS range,
\begin{eqnarray}
g(q^2=0) \sim \frac{\alpha}{2\pi}\chi^2\frac{y_f}{m_f}
\sim 2\times 10^{-6}\ {\rm GeV}^{-1} \ 
\left( \frac{\chi}{10^{-6}}\right)^2 
\left( \frac{y_f\ {\rm eV}}{m_f}\right) 
\,.
\end{eqnarray}   
Interestingly enough, the effective form factor appearing in the   
fractional electric charge~(\ref{eff_charge}) for large photon virtualities, $q^2\gg \mu^2$,  
leads to the fact that, for a scalar $\phi$, the effective Yukawa coupling to the 
proton is suppressed. Therefore, by chosing the paraphoton mass small enough, $\mu\sim$~meV, 
this hidden sector scalar $\phi$ can be a viable candidate for an ALP $0^+$ interpretation
of PVLAS, while nevertheless contributing negligibly to deviations from 
Newtonian gravity in torsion-balance experiments\cite{Dupays:2006dp}.

Finally, let us point out that the required multiple U(1) factors, 
the size of gauge kinetic 
mixing\cite{Dienes:1996zr,Lust:2003ky,Abel:2003ue,Abel:2004rp,Batell:2005wa,Blumenhagen:2006ux},
and suitable matter representations to explain the PVLAS data occur very naturally within the
context of realistic embeddings of the standard model based into string 
theory, in particular in brane world scenarios\cite{Abel:2006qt}.   
 
\section{Conclusions}
  
The evidence for a vacuum magnetic dichroism found by PVLAS has triggered a lot
of theoretical and experimental activities: 
\begin{itemize}
\item Particle interpretations alternative to an axion-like particle interpretation
have been developed, e.g. the minicharged particle interpretation. 
\item Models have been found which evade very strong astrophysical and cosmological  
bounds on such weakly interacting light particles. These models, typically, require even 
more weakly interacting light particles than just the ones introduced for the solution of the PVLAS puzzle, 
a particular example being additional light vector particles (paraphotons).
\item In the upcoming year,
a number of decisive laboratory based tests of the particle interpretation of
the PVLAS anomaly will be done. More generally, these experiments will dig into
previously unconstrained parameter space of the above mentioned models.
\end{itemize} 
Small, high-precision experiments, exploiting high fluxes of low-energy photons and/or
large electromagnetic fields, may give important information about fundamental
particle physics complementary to the one obtainable at high energy colliders!

\section{Acknowledgements}
  
I would like to thank all my collaborators in various aspects of this exciting
field, in particular Steve Abel, Markus Ahlers, Holger Gies, Mark Goodsell, 
Joerg Jaeckel, Ulrich Koetz, 
Valentin V. Khoze, Axel Lindner, Eduard Masso, Raul Rabadan, Javier Redondo, 
Kris Sigurdson, Fuminobu Takahashi, 
and Thomas Tschentscher 
for many discussions and for their support.

\end{document}